\documentclass[conference, letterpaper]{IEEEtran}
\IEEEoverridecommandlockouts
\usepackage{cite}
\usepackage{amsmath,amssymb,amsfonts}
\usepackage{booktabs}
\usepackage{algorithmic}
\usepackage{graphicx}
\usepackage{lipsum}
\usepackage{soul}
\usepackage{subcaption}
\usepackage{textcomp}
\usepackage{xcolor}
\usepackage{listings}
\usepackage{tabularx}
\usepackage{booktabs}
\usepackage{afterpage}
\usepackage[hyphens]{url}
\usepackage{longtable}
\usepackage{array}
\usepackage{multirow}
\usepackage{tabularx}

\usepackage{url}
\usepackage{xspace}

\newcommand{\mypar}[1]{\smallskip\noindent\textbf{#1.}\xspace}
\newcommand\todo[1]{\textcolor{red}{#1}}

\def\BibTeX{{\rm B\kern-.05em{\sc i\kern-.025em b}\kern-.08em
    T\kern-.1667em\lower.7ex\hbox{E}\kern-.125emX}}

\begin{document}

\title{MALET \& Katalina: A Large-Scale Dataset and Static Analysis of macOS Malware}

\author{
\IEEEauthorblockN{Godwin Attigah}
\IEEEauthorblockA{\textit{Independent Researcher}\\
New York, NY, USA \\
godwin@godwinattigah.com}
\and
\IEEEauthorblockN{Obinna Igbe}
\IEEEauthorblockA{\textit{Independent Researcher}\\
New York, NY, USA \\
obinnaigbe@gmail.com}
\thanks{Both authors contributed equally. Author order is lexicographic.}
}

\maketitle

\begin{abstract}
While malware on platforms such as Windows and Android has been extensively studied, the macOS ecosystem remains comparatively understudied. To address this gap, we present MALET, a large-scale public dataset for macOS malware analysis. MALET contains 44{,}804 malicious Mach-O binaries and 44{,}457 samples with zero antivirus detections on VirusTotal. From the undetected set, we further identify 25{,}835 high-confidence benign samples using conservative, heuristic-based filtering. We also introduce Katalina, an open-source static analysis framework that extracts macOS-specific structural features and indicators of potential behavior at scale.

Using MALET, we perform, to the best of our knowledge, the first large-scale static characterization of macOS malware. Our analysis reveals macOS-specific traits, including widespread misuse of security entitlements, abuse of native scripting interfaces, and irregularities in code-signing infrastructure. We find that 93.0\% of malicious samples are unsigned, demonstrating that unsigned code continues to execute in practice through user-mediated approval mechanisms. Among CA-signed malware, 28.4\% (816 samples) were associated with certificates that were later revoked, including samples publicly attributed to DPRK-linked threat actors, as reflected in antivirus labels and threat intelligence reports. We further show substantial revocation latency, with some certificates remaining valid for over 700 days after public detection. Finally, we identify significant detection instability: in 185 structurally identical malicious samples, antivirus engines failed to provide consistent detections for 77.2\% (143 samples). Together, these results highlight persistent blind spots in macOS malware detection and motivate improved approaches to certificate abuse analysis, similarity-based clustering, and platform-specific defense mechanisms. Based on these insights, we outline concrete steps to advance the state of macOS malware analysis and defense.
\end{abstract}


\section{Introduction}
The adoption of macOS has steadily increased, especially within enterprise environments, driven by user preference and perceived security~\cite{apple-in-the-enterprise-2024}. However, this growing popularity also brings heightened attention from cybercriminals, challenging the traditional perception of macOS as inherently secure. Studies show that macOS users often exhibit greater security complacency than users of other platforms, creating vulnerabilities adversaries actively exploit~\cite{topcuoglumacos2024, arce2018malware}. Although Apple has implemented robust defensive technologies like Gatekeeper, Notarization, and Transparency, Consent, and Control (TCC)~\cite{apple_protections}, modern malware authors have proven adept at evading these barriers. Recent campaigns have circumvented Notarization by distributing fake applications through deceptive channels~\cite{patrick_app_notorization}, while others abuse stolen developer certificates to sidestep code-signing validation~\cite{macos-stealers-growing-unit42-2025}. These trends expose critical blind spots in the current macOS security model and emphasize the need for robust macOS malware detection mechanisms.

Despite the growing threat landscape, macOS remains underrepresented in academic security research compared to the extensive work on Windows (BODMAS \cite{yang2021bodmas}), Android (Drebin\cite{lindorfer2014andrubis}), and Linux (Understanding Linux Malware \cite{cozzi2018understanding}) malware. This gap is partly due to the unique challenges of conducting dynamic analysis on macOS. Sandboxing restrictions and the transition from Intel (x86\_64) to Apple Silicon (ARM) complicate binary execution, often requiring translation layers or dual-architecture setups that are difficult to scale. In addition, dynamic analysis typically demands access to genuine Apple hardware since virtualization on non-Apple devices is restricted by licensing, and emulation remains unreliable. 
These challenges make static analysis a compelling alternative, capable of scaling to thousands of samples without execution risk. Furthermore, the lack of large publicly available and well-labeled datasets for macOS, unlike those available for other platforms such as EMBER\cite{anderson2018ember} (Windows), SOREL 20M\cite{harang2020sorel} (Windows), and ANDRUBIS\cite{lindorfer2014andrubis} (Android), hinders reproducibility and systematic evaluation in macOS-focused research.


To address this gap, we introduce \textbf{MALET}, the largest publicly available dataset of macOS Mach-O binaries, and \textbf{Katalina}, an open-source static analysis tool that extracts structural features and static indicators of potential behavior at scale. These indicators include, but are not limited to, entitlements, signing metadata, embedded scripts, and linked libraries. Together, MALET and Katalina provide a reproducible foundation for systematic macOS malware analysis.

\noindent Using MALET, we conduct, to the best of our knowledge, the first large-scale static characterization of macOS malware and report the following findings:

\begin{itemize}
    \item \textbf{Unsigned prevalence}: 93.0\% of malicious samples are unsigned, indicating that unsigned binaries continue to execute in practice under user-mediated approval mechanisms.
    \item \textbf{Certificate exposure}: 28.4\% of CA-signed malicious samples were associated with certificates that were later revoked, including cases publicly attributed to DPRK-linked threat actors, with some certificates remaining valid for over 700 days.
    \item \textbf{Scripting interface misuse}: Malware frequently abuses \texttt{osascript} for credential access and data exfiltration, analogous to PowerShell abuse on Windows.
    \item \textbf{Labeling inconsistencies}: In a set of 185 structurally identical malicious samples, antivirus engines failed to raise consistent detections for 77.2\% (143 samples), highlighting substantial labeling noise.
    \item \textbf{Low packer prevalence}: Only 1.36\% of malicious samples in MALET used commodity packers (exclusively UPX).
\end{itemize}

These results reveal persistent blind spots in macOS malware detection and emphasize the need for platform-specific detection strategies and attribution research.

This paper makes the following contributions:

\begin{itemize}
    \item \textbf{Dataset}: We introduce MALET, comprising 44{,}804 malicious and 44{,}457 undetected Mach-O binaries (i.e., with zero AV detections), annotated with AVClass2 family and class labels. The undetected subset includes 25{,}835 high-confidence benign samples identified via conservative, heuristic-based filtering.
    \item \textbf{Tooling}: We develop Katalina, an open-source static analysis framework that extracts macOS-specific structural features and indicators of potential behavior, including entitlements, signing metadata, embedded scripts, and dependency relationships.
    \item \textbf{Empirical analysis}: We present a comprehensive measurement study of macOS malware, identifying characteristic patterns such as entitlement misuse, scripting interface abuse, code-signing anomalies, and detection instability across structurally identical samples.
\end{itemize}

You can find Katalina at \todo{\url{https://github.com/appsworld/katalina}} and MALET at \todo{\url{https://zenodo.org/records/17877273}}.


\section{Background and Motivation}
\label{sec:background}
Understanding macOS malware requires familiarity with Apple’s binary format, modern system defenses, and the practical limits of static and dynamic analysis. This section outlines the Mach-O file structure, key security mechanisms such as notarization and SIP, and challenges that shape macOS malware research.

\subsection{Mach-O Binary Format}
The Mach-O (Mach Object) file format is the native executable format used by macOS for executables, dynamic libraries, dynamic linkers, core dumps, and object files \cite{levin2012mac}. Like PE (Portable Executable) files, which serve as the primary executable format for Windows, Mach-O provides the foundation for program execution on Apple platforms.

\subsubsection{File Structure Overview}
\begin{figure}[t]
\centering
\includegraphics[width=0.8\columnwidth, keepaspectratio=true]{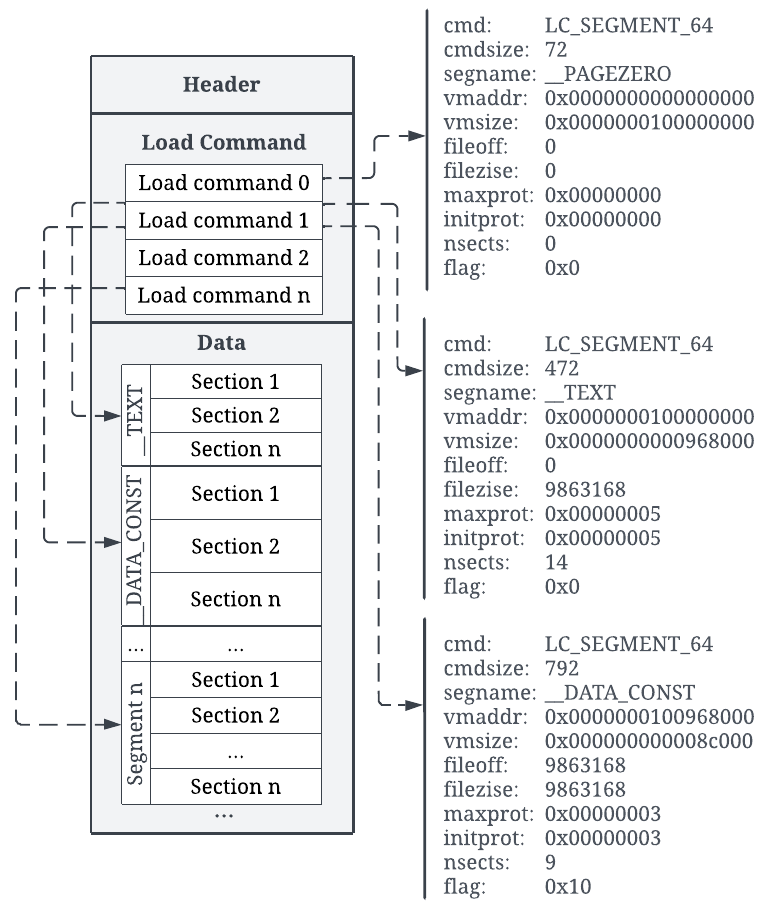}
\caption{The Mach-O file format with example values.}
\label{fig_macho}
\end{figure}

A Mach-O binary consists of three main parts: a header, load commands, and data segments. Figure~\ref{fig_macho} illustrates this structure. The header identifies the file as a Mach-O binary and provides basic information about its architecture and organization. Load commands specify how to set up the program in memory, and data segments contain the actual code and data of the program.

Multiple Mach-O binaries can be combined into a universal (or FAT) binary, allowing applications to run on different architectures (for example, Intel x86\_64 and Apple ARM64)~\cite{apple2020universalbinary}. A Universal binary begins with a header that identifies the file type, architecture count, and metadata for each contained Mach-O binary~\cite{llvm2025machoheader}. Following Apple's 2020 transition to ARM-based systems~\cite{apple2020transition}, developers increasingly adopted Universal binaries. This complicates malware analysis, as each architecture variant must be examined separately. In Section~\ref{subsec:katalina}, we explain how Universal Binaries now require malware analysis tools to explicitly choose and extract a platform-specific architecture for analysis, which complicates analysis workflows when malware authors embed distinct traits for each platform.

\subsubsection{Header Structure}
The Mach-O header contains essential metadata required for interpreting the binary. It begins with the \texttt{magic number}, a unique identifier that marks the file as a Mach-O binary. This is followed by the \texttt{CPU type} and \texttt{subtype}, which specify the target architecture, such as \texttt{x86\_64} or \texttt{ARM64}. The \texttt{file type} field indicates the role of the file, whether it is an object file, executable, dynamic library, bundle or core dump. The header also includes the \texttt{number} and \texttt{size} of load commands, which define how many instructions follow the header~\cite{apple_macho_loader_gh}. Finally, the \texttt{flags} field specifies optional features, such as \texttt{MH\_NOUNDEFS}, which indicates the object file had no undefined references at build time. These flags are conceptually similar to the characteristics field in PE headers, helping the loader determine how to handle the binary~\cite{apple_macho_loader_gh, pietrek1994peering}.

\subsubsection{Load Commands}
The load command table follows the Mach-O header and provides instructions for loading various file parts into memory. Each load command has a type (\texttt{cmd}) and size (\texttt{cmdsize}). Over 40 supported load commands are in the format \cite{apple_macho_loader_gh}. Among these commands, the segment commands (\texttt{LC\_SEGMENT} and \texttt{LC\_SEGMENT64}) are critical. They define memory segments to be mapped into a process's address space and are analogous to section tables in PE files. 

Key fields in these commands include \textbf{vmaddr} (starting virtual memory address), \textbf{vmsize} (virtual memory size), \textbf{fileoff} (file offset of segment data), \textbf{filesize} (data size in file), \textbf{initprot} (initial protection flags), and \textbf{maxprot} (maximum protection allowed).

Common segment commands include \texttt{\_\_PAGEZERO} (a trap for NULL pointers), \texttt{\_\_TEXT} (executable code), \texttt{\_\_DATA} (writable non-constant data), \texttt{\_\_DATA\_CONST} (read-only global constants), and \texttt{\_\_LINKEDIT} (symbol tables and linking information). These serve functions similar to PE counterparts like .text, .data, and .rdata sections.

The segments are divided into sections, each with a unique name and purpose. For example, the \texttt{\_\_text} section typically contains the executable code, while the \texttt{\_\_data} section contains initialized global and static variables. Other common sections include \texttt{\_\_cstring}, \texttt{\_\_const}, \texttt{\_\_got}, and \texttt{\_\_la\_symbol\_ptr}.

Another important command is \verb+LC_LOAD_DYLIB+, which specifies external dynamic libraries required during execution. Dynamic libraries (\emph{.dylib} files) in macOS serve a similar role as DLLs on Windows. The \verb+LC_LOAD_DYLIB+ command stores the library name, timestamp, current version, and compatibility version~\cite{osmach}. At runtime, the dynamic linker uses the library name to locate the corresponding file, and the compatibility version ensures that the binary links only against compatible library versions.

\subsection{Application Packaging in macOS}

While the previous section focused on the Mach-O binary structure, binaries often arrive in packaged form rather than as raw executables. macOS supports several packaging formats that encapsulate Mach-O binaries and supporting resources~\cite{apple_packaging_mac_software}. These include \texttt{Application Bundles (.app)}, which are directory structures containing resources and the primary executable in \texttt{Contents/MacOS/}; \texttt{Disk Images (.dmg)}, which serve as mountable volumes for distributing apps or installers; \texttt{Installer Packages (.pkg)}, which bundle files, scripts, and installation metadata; and \texttt{Compressed Archives (.zip)}, which provide compressed containers for delivery.
Understanding these formats is essential for malware analysis, as threats are often embedded within these containers. The Archive Extraction step in Phase 2 of Figure~\ref{fig:malet_processing} describes how we extract Mach-O binaries and process each packaging type during dataset construction.

\subsection{Code Signing and Certificate Trust}
Since macOS 10.15, Apple requires developers to sign and notarize applications distributed outside the App Store using a Developer ID certificate to comply with Gatekeeper policies~\cite{apple2024platformsecurity}. A valid signature ensures the binary has not been modified and suppresses user prompts at execution. Malware authors exploit this trust model by acquiring valid certificates, reusing stolen ones, or distributing unsigned binaries. In Section~\ref{mypar:certificates-macos-malware}, we analyze how attackers abuse the signing process, including cases involving fraudulently obtained certificates.
Apple's certificate hierarchy includes the Apple Root CA, the Developer ID Certification Authority (intermediate), and Developer ID Application or Installer certificates (end-entity). Developers sign binaries with their private key, and macOS verifies signatures through the public key chain anchored at Apple.
Each Developer ID includes a unique Team ID (e.g., \texttt{BJ4HAAB9B3} for Zoom) on all associated certificates~\cite{zoom_kext_changes}. We use Team IDs to identify reused or revoked certificates and link malware to specific campaigns in Section~\ref{mypar:certificates-macos-malware}.

\subsection{Entitlements and Security Controls}
Entitlements are a key security feature in macOS that controls the capabilities of an application at runtime. Similar to Android permissions but less visible to users, entitlements are specified in property list (plist) files embedded within the binary and are tightly coupled with code signing. They grant access to protected resources such as cameras, location services, or system APIs.
Mandatory Access Controls (MACs) in macOS require code signing to enable system-protected entitlements~\cite{apple2024platformsecurity}. For example, applications that access sensitive resources such as the camera or location services must include entitlements like com.apple.security.device.camera and com.apple.security.personal-information.location. Other entitlements, such as com.apple.security.cs.allow-jit and com.apple.security.get-task-allow, govern runtime behavior by enabling just-in-time code execution and debugging. In Section~\ref{sec:characterizing_macos}, we analyze how malware abuses these entitlements to gain privileged access and bypass system protections.

\subsection{Malware Mitigations in macOS}
The macOS operating system incorporates several defense mechanisms to protect against malicious software~\cite{apple_protections}:
\begin{itemize} \item \textbf{Notarization}: A feature that confirms software authenticity and safety. Before distribution outside the App Store, applications must be signed with a valid developer certificate and pass Apple's malware checks~\cite{apple_protections}.
\item \textbf{Gatekeeper}: Gatekeeper protects users by preventing the execution of potentially harmful applications downloaded to the Internet. It tags downloaded apps with a quarantine attribute, triggering verification of the app's signature and notarization. Apps that fail this verification are blocked from execution~\cite{applegatekeeper2021april}.

\item \textbf{XProtect}: A built-in antivirus that uses YARA signatures to detect and block known malware from executing~\cite{apple_protections}.

\item \textbf{TCC}: The Transparency, Consent, and Control (TCC) feature allows users to control, monitor, and revoke app permissions for sensitive data or device functions such as camera and microphone~\cite{apple_tcc}.

\item \textbf{SIP}: The System Integrity Protection (SIP) protects critical system files and processes from unauthorized modifications without explicit permission from the system~\cite{apple_sip}.
\end{itemize}

Despite these protections, malware authors find ways to circumvent these mechanisms. For example, attackers exploit framework vulnerabilities like those in \texttt{FileProvider} (which manages file access between applications) to bypass TCC restrictions~\cite{jamf2023tcc} or abuse system services such as \texttt{ storagekitd} (a daemon that manages storage operations) to evade SIP~\cite{microsoft2025cve44243}.

\subsection{Challenges in macOS Malware Analysis}
macOS malware analysis introduces challenges that are not found on other platforms. Apple's shift from Intel to ARM-based Apple Silicon forces researchers to use Apple-specific hardware and account for architecture-specific behavior, which limits scalabale dynamic analysis~\cite{apple2020universalbinary}. SIP and TCC further restrict access to system components, making runtime inspection difficult. These constraints increase the value of static analysis, especially approaches that do not depend on platform-specific execution. Static analysis enables researchers to examine binaries, extract resources, and infer behavior without running the malware~\cite{stokes2019reverse}.

However, effective static analysis requires a deep understanding of macOS-specific constructs such as the Mach-O file format, entitlements, property lists, and code signing, all explored in this section. Compounding the difficulty is the lack of large-scale, publicly available macOS malware datasets. Unlike Windows and Android, which benefit from extensive corpora and standardized tooling, macOS research has been limited by small sample sets, inconsistent labeling, and fragmented feature extraction methods. This gap motivates the development of the MALET dataset, which we introduce in Section~\ref{sec:malet_dataset}.

\section{Data Collection and Processing}
\label{sec:data_collection}

\begin{figure*}[t] 
\centering
\includegraphics[width=0.9\textwidth]{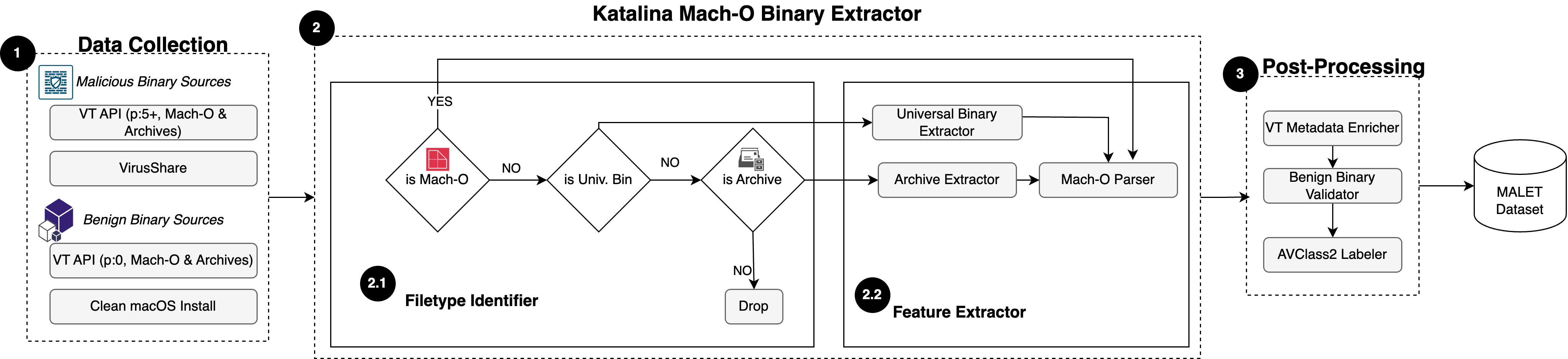} \caption{Overview of the MALET dataset construction pipeline. Processing proceeds in three phases: (1) Data Collection, in which macOS binaries are collected from VirusTotal and VirusShare using detection-based criteria ($\geq$5 detections for malware, 0 detections for undetected software), as well as from a clean macOS installation; (2) Binary Extraction and Parsing, where valid Mach-O files are identified, extracted from universal binaries and archive formats when applicable, and processed by Katalina to extract static features; and (3) Post-Processing, which enriches malware samples with VirusTotal metadata, assigns standardized family and class labels via AVClass2, and annotates undetected samples with benign-confidence signals derived from trust metadata. The MALET dataset contains malicious samples and undetected software with benign-confidence annotations; detection absence alone is not treated as benign ground truth.} 
\label{fig:malet_processing}
\vspace{-0.5cm}
\end{figure*}

To better understand the macOS malware landscape, we curated a large-scale dataset of Mach-O binaries comprising malicious samples and undetected software annotated with benign- confidence signals. As shown in Figure~\ref{fig:malet_processing}, dataset construction proceeds in three phases: Phase~1- data collection, Phase~2- Mach-O parsing and feature extraction, and Phase~3- post-processing for validation and enrichment.

\subsection{Data Collection} 
Phase~1 constructs a macOS malware and non-malware corpus of Mach-O binaries. Malicious samples are obtained from VirusTotal and VirusShare, while non-malicious samples are derived from fresh macOS installations and VirusTotal samples with zero detections. Because undetected samples do not imply benignity, we apply provenance-based heuristics such as code-signing identity, notarization status, and vendor reputation to identify a high-confidence benign subset.

\textbf{Malicious Binaries Acquisition.} We obtained malicious samples from two well-established repositories: \textit{VirusTotal} \cite{vt_search_home} and \textit{VirusShare} \cite{viirus_share_home}, which are commonly used in prior malware research for initial data collection \cite{avclass2}.

\textit{VirusTotal API (p:5+)}: We searched VirusTotal for Mach-O binaries and related artifacts, including disk images, application bundles, and package installers. We filtered results using the p:5+ modifier to select files flagged by five or more antivirus engines. Prior work~\cite{avclass2, av_label_changes} shows that this method improves precision while preserving the coverage of real-world macOS malware.

\textit{VirusShare Archives}: We downloaded generic malware archives from VirusShare \cite{virus_share_corpus} using their bulk corpus download option. Unlike Windows and Linux, VirusShare doesn't offer macOS-specific archives, so we processed these general collections to extract potential macOS samples. We used Phase 2.1  of the Katalina pipeline to identify and extract Mach-O binaries from these archives for further analysis.

\textbf{Non-malicious Binary Acquisition.} To characterize non-malicious macOS software, we collected an \emph{undetected} set consisting of binaries with zero antivirus detections on VirusTotal at collection time. Because the absence of detections does not guarantee benignity, we applied additional heuristics to annotate a subset of undetected samples as \emph{high-confidence benign}. These heuristics leverage software trust and provenance signals, including Apple notarization, CA-based code signing, and trusted developer identities (see Phase~3 for details).

\textit{Clean macOS Installation}: We additionally extracted binaries from a freshly installed macOS environment configured with macOS 12.6 (21G115). To identify Mach-O binaries, we traversed the filesystem from the root directory and used the \emph{file} command to confirm each file as a Mach-O binary or extractable archive. These system binaries were annotated using the same benign-confidence logic and serve as reference samples drawn from a controlled macOS installation environment.

\textit{VirusTotal API (p:0)}: To expand beyond system-provided binaries, we queried VirusTotal for macOS samples with zero antivirus detections (p:0) using filters analogous to those used for malware discovery, but with detection criteria inverted. This allowed us to collect third-party macOS applications from the broader ecosystem. Because the absence of detections does not imply benignity, all candidates were annotated using the same benign- confidence validation pipeline described in Phase~3.

\subsection{Katalina Tool: Mach-O Binary Extractor} 
\label{subsec:katalina}
The Katalina tool in Phase 2 is our custom-built, cross-platform Golang utility for extracting and parsing Mach-O binaries from various macOS binary formats. It operates in two phases: File Identification to identify parsable samples and Feature Extraction to retrieve and normalize Mach-O features.

\textbf{File Identification and Isolation Logic (Phase 2.1).} 
The extraction pipeline begins by classifying each input file into one of three categories: (i) native Mach-O binaries, which are parsed directly; (ii) universal (FAT) binaries containing multiple architecture-specific payloads, each of which is extracted and analyzed independently by Katalina; and (iii) archives or container formats (e.g., DMG, PKG, ZIP), which are unpacked using \texttt{7z} prior to analysis.

To isolate the executable binary (that is, the actual candidate executable), we retrieve the  $CFBundleExecutable$ key from the binary's \texttt{Info.plist}.  In the Apple ecosystem the $CFBundleExecutable$ is used to locate the executable or dylib that binary expects macOS to load on execution\cite{apple_cfbundle}. We then search and retrieve the candidate binary ensuring that its name matches the retrieved value.

\textbf{Feature Extraction: Phase 2.2} After identifying Mach-O types, Katalina extracts structural features by reading headers, segments, and sections directly from load commands, preserving all metadata. A custom XML parser retrieves entitlements from the code signing section, supporting DER-encoded and plist formats. The handling of certificates includes extracting the metadata from the issuer, validating the signing chains, and performing OCSP revocation checks. Katalina also extracts Objective-C class and method names when present, normalizes and sorts them, and computes SHA-256 hashes to generate consistent fingerprints. To support extraction of instruction-level features (e.g., opcode distributions), we implemented a lightweight disassembler in Katalina. This enables profiling of instruction usage patterns across each binary. String-based features such as URLs, domain names, file paths, and artifacts related to shell usage or script execution (e.g., references to \texttt{/bin/bash}, \texttt{/bin/sh}, launch daemon paths, and TCC database access) are extracted using regular expressions based on known macOS persistence and privilege escalation techniques. We also compute entropy distributions across sections and strings to detect packing and obfuscation.

\subsection{Post-Processing}
\label{subsec:post-processing}

Following feature extraction, we persist all extracted features and metadata in a long-term research database for downstream analysis. Phase~3 applies post-processing to enrich malware samples with VirusTotal (VT) metadata and to annotate undetected samples with benign-confidence flags derived from software trust signals. As demonstrated in Section~\ref{mypar:uuid-collisions}, we observe instances where samples with zero detections are later associated with malicious behavior, motivating the need for conservative validation logic rather than treating undetected samples as ground truth benign.

\textbf{VirusTotal Metadata Enricher: Phase 3.1} We query the VirusTotal API using SHA-256 hashes to retrieve antivirus verdicts, first-seen timestamps, and sample tags for malware samples. These metadata are required inputs to the AVClass2 labeling pipeline and enable temporal analysis across families and campaigns. Malware samples obtained from VirusShare that were not present in VirusTotal were excluded from MALET due to lack of labeling metadata.

\textbf{Benign Binary Validator: Phase 3.2} We implement a tiered validation pipeline to reduce the risk of misclassifying undetected malware as benign. Rather than constructing a separate dataset, we annotate undetected samples with a \emph{high-confidence benign} flag derived from trust and provenance signals. Global exclusion rules eliminate samples associated with revoked certificates, known abuse-linked developer identities, or developer tooling namespaces that are frequently abused for malware staging. Remaining candidates are evaluated against multiple trust tiers, including (i) CA-signed and Apple-notarized binaries, (ii) binaries attributed to widely trusted software publishers, and (iii) Apple system components identified via bundle identifiers. A sample is annotated as high-confidence benign if it satisfies at least one trust tier after passing all exclusions. This approach prioritizes precision over recall and treats benign-confidence as a trust signal rather than ground truth.

\textbf{AVClass2 Labeler: Phase 3.3} We employ AVClass2~\cite{avclass2} to generate normalized malware family and class labels from heterogeneous antivirus detection strings. AVClass2 applies rule-based parsing and majority voting across engines to assign the most likely family and behavior class. These labels standardize vendor-specific naming and support longitudinal analysis across malware campaigns. We discuss limitations of AVClass2 for macOS family resolution in Section~\ref{section:avclass2}.

\begin{table}[!t]
\caption{MALET Dataset Summary}
\centering
\begin{tabular}{lrrr}
\toprule
 & \textbf{Malicious} & \textbf{Undetected} & \textbf{Total} \\
\midrule
\textbf{Total samples} & 44{,}804 & 44{,}457 & 89{,}255 \\
\midrule
Universal binary       & 15{,}659 & 27{,}170  & 42{,}829 \\
\midrule
\multicolumn{4}{l}{\textbf{Signature}} \\
\quad CA-signed        & 2{,}869  & 37{,}772 & 40{,}641 \\
\quad Unsigned         & 41{,}655 & 6{,}691  & 48{,}346 \\
\quad Revoked          & 816      & 559      & 1{,}375  \\
\midrule
\multicolumn{4}{l}{\textbf{Architecture}} \\
\quad Amd64 (x86\_64)  & 30{,}582 & 23{,}218 & 53{,}796 \\
\quad AARCH64 (ARM64)  & 9{,}419  & 20{,}385 & 29{,}802 \\
\quad i386             & 4{,}598  & 268      & 4{,}866  \\
\quad ARM              & 46       & 494      & 540      \\
\quad PowerPC          & 155      & 74       & 229      \\
\quad PowerPC 64       & 4        & 4        & 8        \\
\quad ARM64\_32        & 0        & 14       & 14       \\
\bottomrule
\end{tabular}
\caption*{\footnotesize All counts are distinct binaries (unique SHA-256 hashes).}
\label{tab:malet_summary}
\end{table}

\begin{table}[!t]
\centering
\caption{Dataset Distribution by Label and Class}
\label{tab:malware-distribution}
\begin{tabular}{llrr}
\toprule
\textbf{Label} & \textbf{Class} & \textbf{Count} & \textbf{Contextual Distribution} \\
\midrule
undetected & N/A & 44,457 & 49.8\% of total \\
\quad high-conf benign & & 25,835 & 29.0\% of total \\
\midrule
malicious & stealer & 14,712 & 32.8\% of malware \\
malicious & ransomware & 13,204 & 29.5\% of malware \\
malicious & backdoor & 6,477 & 14.5\% of malware \\
malicious & grayware & 3,479 & 7.8\% of malware \\
malicious & miner & 3,200 & 7.1\% of malware \\
malicious & unknown & 1,700 & 3.8\% of malware \\
malicious & hacktool & 1,216 & 2.7\% of malware \\
malicious & keylogger & 478 & 1.1\% of malware \\
malicious & downloader & 261 & 0.6\% of malware \\
\bottomrule
\end{tabular}
\end{table}

\begin{table}[t]
\centering
\caption{Top malware families in MALET (distinct malicious samples).}
\label{tab:malware-family-distribution}
\begin{tabular}{l l r r}
\toprule
\textbf{Family} & \textbf{Class} & \textbf{Samples} & \textbf{\% of malware} \\
\midrule
amos      & stealer    & 13{,}950 & 31.1\% \\
evilquest & ransomware & 12{,}532 & 28.0\% \\
flashback & backdoor   &  3{,}326 &  7.4\% \\
unknown   & miner      &  3{,}058 &  6.8\% \\
unknown   & unknown    &  1{,}677 &  3.7\% \\
adload    & grayware   &  1{,}315 &  2.9\% \\
unknown   & backdoor   &    695   &  1.6\% \\
keranger  & ransomware &    652   &  1.5\% \\
unknown   & stealer    &    648   &  1.4\% \\
unknown   & grayware   &    580   &  1.3\% \\
\bottomrule
\end{tabular}
\end{table}

\subsection{The MALET Dataset} 
\label{sec:malet_dataset}

After completing the full processing pipeline and discarding unparsable samples, we assembled the MALET dataset containing \textbf{89{,}255} unique macOS binaries: \textbf{44{,}804} labeled malicious (50.2\%) and \textbf{44{,}457} samples with zero VirusTotal detections annotated as undetected (49.8\%). Within the undetected set, we annotate \textbf{25{,}835} samples (29.0\% of all unique samples) with a \emph{high-confidence benign} flag using conservative trust-based heuristics (Section~\ref{subsec:post-processing}).

Table~\ref{tab:malet_summary} summarizes key dataset characteristics. Universal Mach-O binaries constitute 42{,}829 samples (48.0\%) of the corpus and are significantly more prevalent among undetected binaries (61.1\%) than malicious ones (35.0\%). Most malicious binaries are unsigned (93.0\%), while the majority of undetected samples are CA-signed (84.9\%). Certificate revocation, determined using the \texttt{IsCertificateRevoked} feature, appears in 1{,}375 samples (816 malicious, 559 undetected); among malicious binaries this corresponds to 28.4\% of CA-signed samples, compared to only 1.5\% among undetected ones.

\begin{figure*}[t]
    \centering
    \begin{subfigure}{0.48\textwidth}
        \centering
        \includegraphics[width=\linewidth, keepaspectratio]{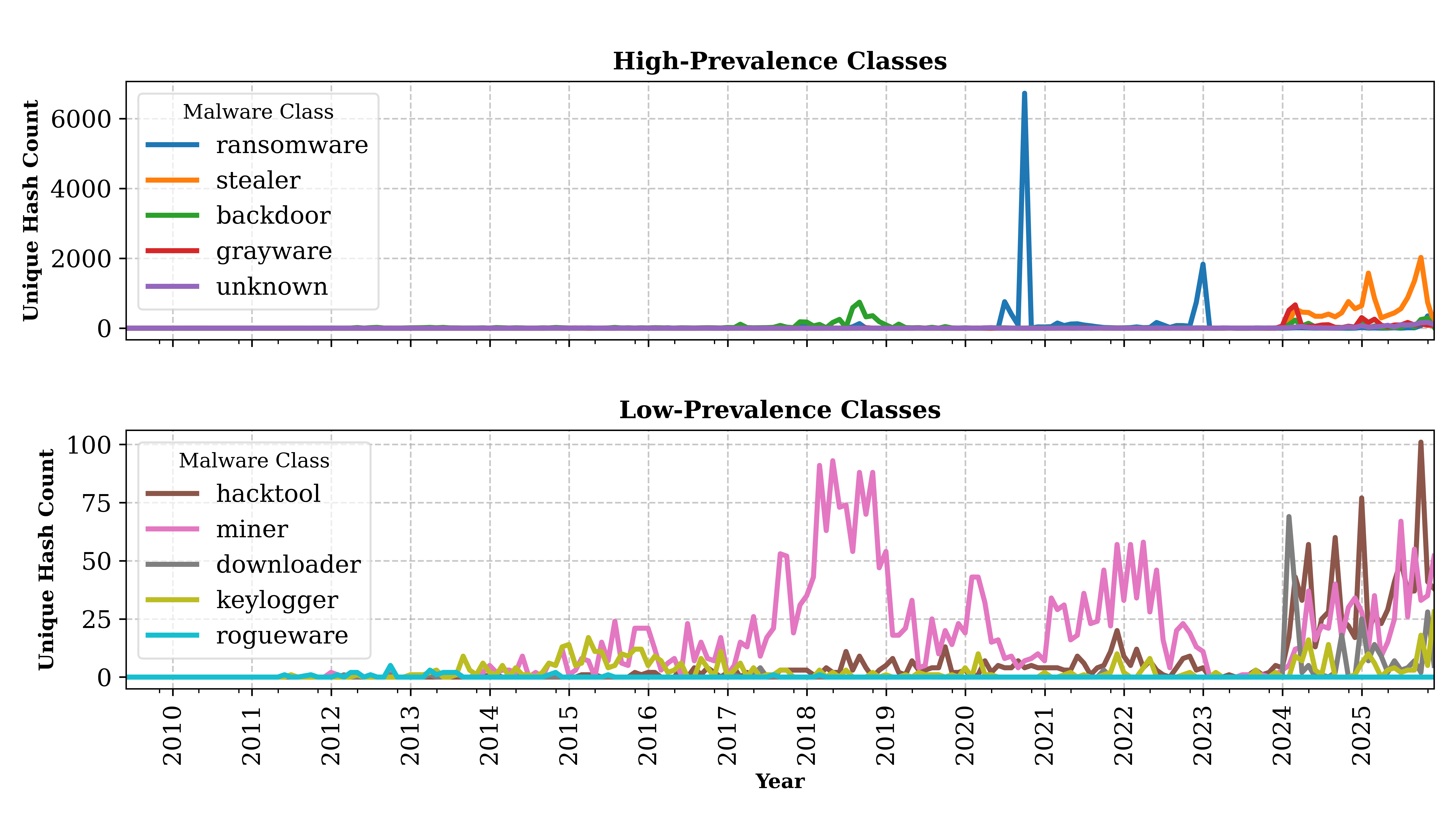}
        \caption{Top ten malware classes.}
        \label{fig:sample_temporal_distribution}
    \end{subfigure}
    \hfill
    \begin{subfigure}{0.48\textwidth}
        \centering
        \includegraphics[width=\linewidth, keepaspectratio]{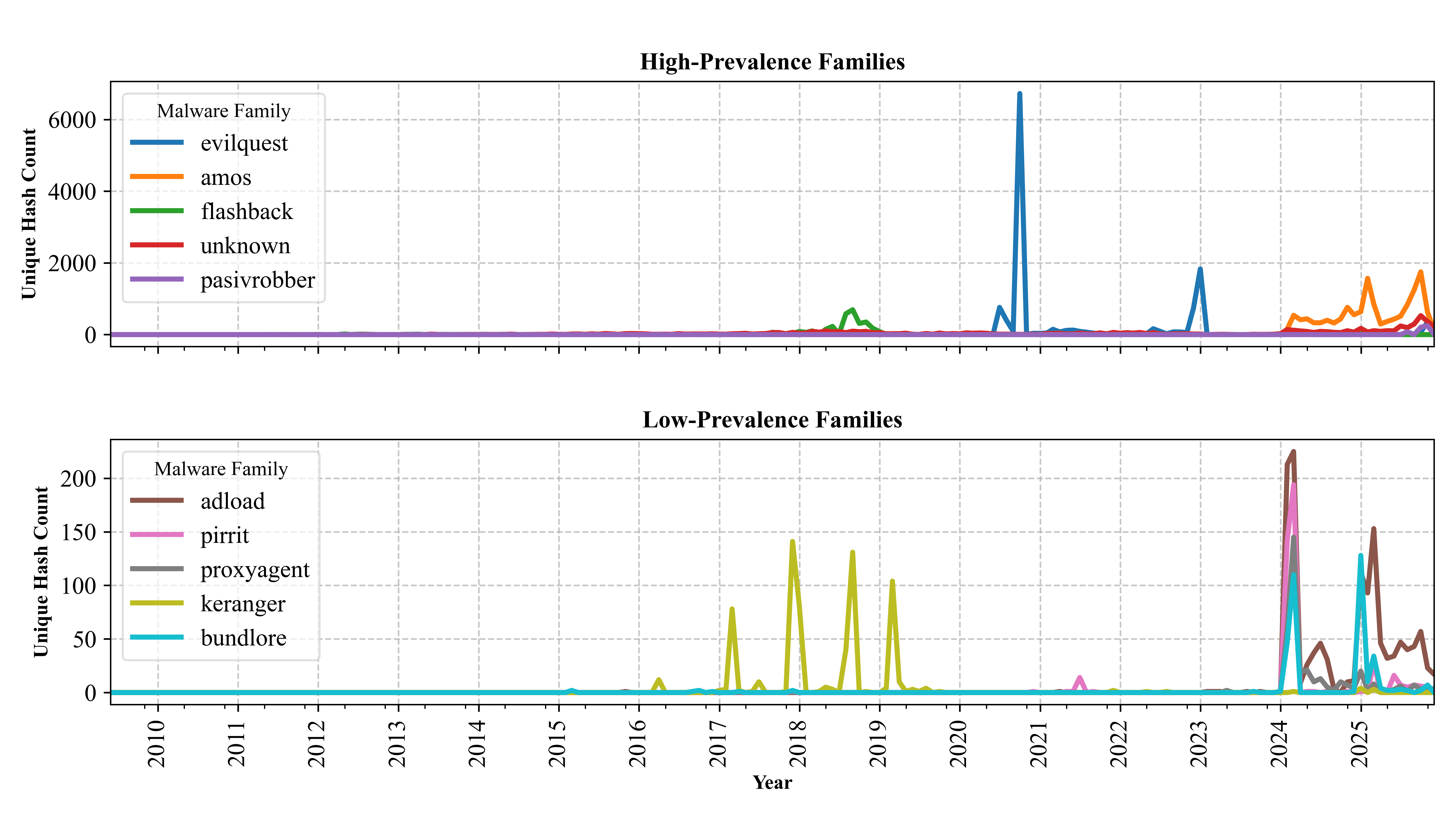}
        \caption{Top ten malware families.}
        \label{fig:family_temporal_distribution}
    \end{subfigure}
    \caption{Temporal distribution of the top ten malware classes (left) and families (right) in MALET, each split by prevalence to preserve visibility across orders of magnitude.}
    \label{fig:temporal_classes_families}
\end{figure*}

Using AVClass2, we mapped AV engine labels to standardized malware class and family labels, identifying 17 distinct malware classes and 212 unique families. Samples without clear classifications were labeled as \textit{unknown}. Table~\ref{tab:malware-distribution} summarizes the class distribution, with \textit{stealer} (32.8\% of malicious samples) and \textit{ransomware} (29.5\%) being the most prevalent classes, followed by \textit{backdoor}, \textit{grayware}, and \textit{miner}. Figure~\ref{fig:sample_temporal_distribution} illustrates the temporal distribution of the top malware classes. We discuss further details of this classification process in Section~\ref{section:classification-macos-malware}. At the family level, Table~\ref{tab:malware-family-distribution} shows that a small number of families account for a disproportionate share of activity, with \textit{amos} and \textit{evilquest} together comprising over half of all malicious samples. In contrast, the
remaining families form a long-tailed distribution with substantially lower prevalence.
Figure~\ref{fig:family_temporal_distribution} highlights distinct lifecycle patterns across families,
including rapid growth trajectories for newer families and gradual decline among older ones,
reflecting the dynamic and campaign-driven nature of the macOS malware ecosystem.

\begin{figure}[t]
    \centering
\includegraphics[width=0.49\textwidth, keepaspectratio]{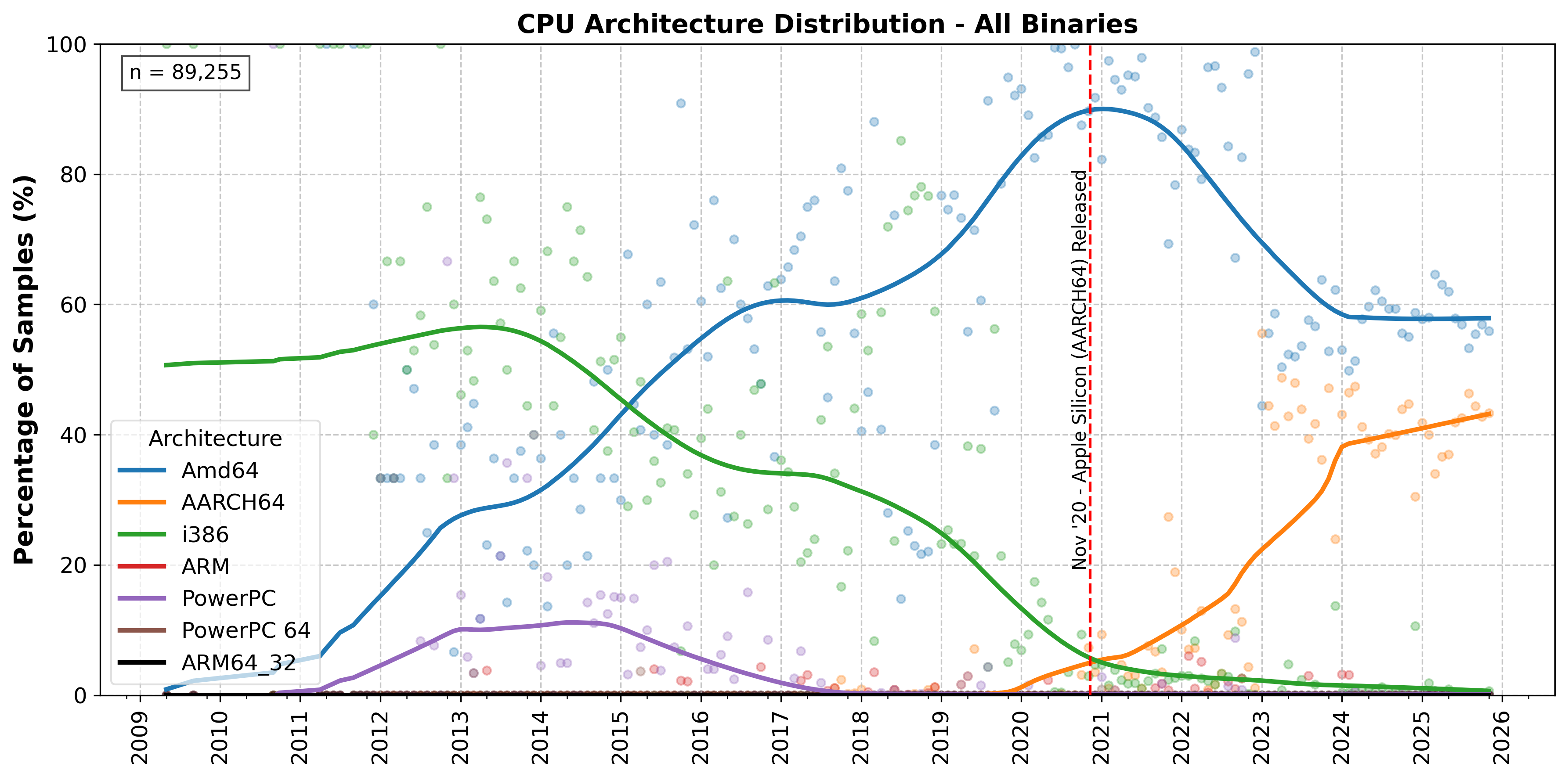}
        \caption{CPU Architecture distribution over time (malicious + undetected samples)}
        \label{fig:cpu_architecture_distribution}
\vspace{-0.5cm}        
\end{figure}

The dataset spans seven CPU architectures, with x86\_64 (53{,}796 samples; 60.3\%) and ARM64 (29{,}802; 33.4\%) accounting for the majority of binaries. The remaining architectures include i386 (5.5\%), ARM (0.6\%), PowerPC (0.3\%), and a small number of PowerPC~64 and ARM64\_32 samples, together contributing less than 1\% of the corpus. Figure~\ref{fig:cpu_architecture_distribution} shows the temporal evolution of these architectures and highlights the rise of ARM64 following Apple’s transition to Apple Silicon~\cite{apple2020transition}.

\section{Characterizing macOS Malware: Insights from the Dataset}
\label{sec:characterizing_macos}

This section presents a set of representative behaviors identified in macOS malware from the MALET dataset. Rather than attempting an exhaustive taxonomy, we highlight behaviors characteristic of macOS threats that have not been explored in prior literature or are particularly salient in understanding this ecosystem. Where relevant, we contextualize these behaviors with quantitative metrics to assess their operational frequency.

Our methodology combines structured queries over the MALET corpus with manual reverse engineering of selected samples. The following subsections analyze key behavioral traits such as misuse of entitlements, abuse of code-signing certificates, interaction with scripting interfaces, and packing in macOS.

\subsection{Entitlements in macOS Malware} \label{mypar:entitlements-macos-malware}
Entitlements in the Apple ecosystem are akin to Android permissions; they grant binaries privileged capabilities such as access to private APIs, memory manipulation, or turning off sandbox constraints. However, Apple entitlements are tightly coupled with code signing and lack user visibility \cite{appleEntitlements}, making them a subtler but potent attack surface. Although Android permission abuse has been widely studied \cite{felt2012android, lin2012expectation, au2012pscout}, similar analysis in macOS remains limited. Android research shows that permissions shape user trust and app behavior \cite{spreitzenbarth2013mobile}, but macOS lacks comparable behavioral insights.

Analysis of recent CVE-2024-44243 demonstrates how attackers exploit the \texttt{storagekitd} daemon to bypass System Integrity Protection through malicious kernel extensions \cite{microsoft2025cve44243}. The vulnerability enables attackers to load unauthorized kernel code, effectively disabling core macOS security measures and expanding the attack surface for additional exploits.

We identified 3,290 unique entitlements across the MALET dataset, 65 of which appeared in exclusively in malicious samples. These entitlements are extracted by Katalina through parsing XML and DER-encoded plist sections embedded in the code signature. Our parser processes the CMS blob and extracts the embedded entitlement dictionary. Although the tool supports Intel and ARM binaries, malformed or obfuscated signatures may evade extraction. Among samples containing at least one entitlement, malicious binaries contained an average of 1.6 entitlements compared to 7.7 in undetected binaries. Despite containing fewer entitlements overall, malware samples preferred certain high-privilege entitlements that enable system-level access.

\paragraph{Analyzing Entitlement Patterns in Malware}

\begin{figure}[t]
    \centering
    \includegraphics[width=0.48\textwidth]{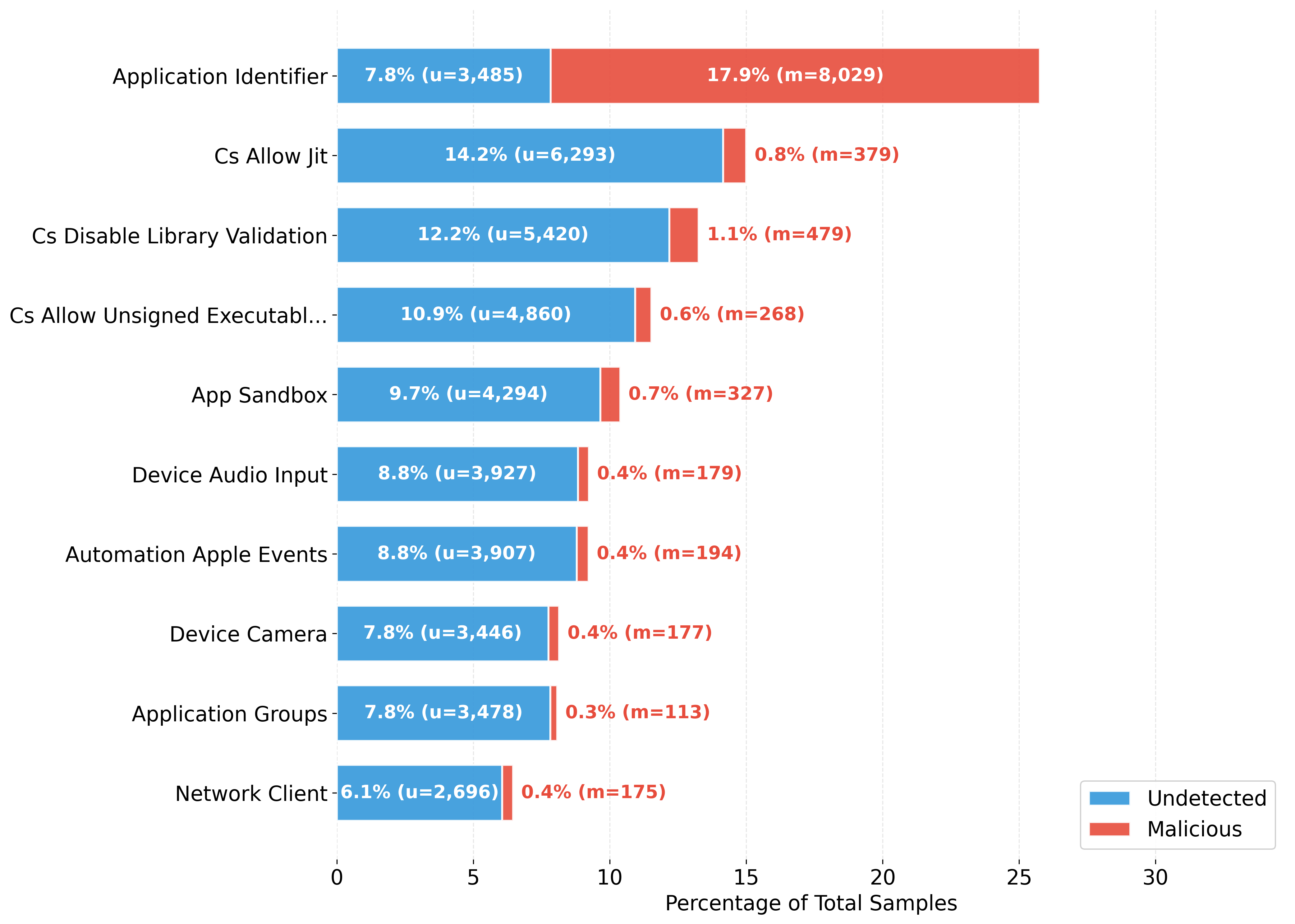}
    \caption{Top 10 entitlements found in macOS binaries. Blue bars represent undetected samples (u), while red bars represent malicious samples (m). Percentages are based on the combined total of all samples, with sample counts shown for each category. }
    \label{fig:entitlement_mitre_matrix}
\vspace{-0.5cm}
\end{figure}

We categorized entitlements by functional purpose and security impact. We found that malware authors frequently target entitlements related to four primary objectives: evading defenses, accessing sensitive data, maintaining persistence, and  bypassing sandboxing restrictions. Figure~\ref{fig:entitlement_mitre_matrix} shows the top ten entitlements in benign and malicious samples.

Several entitlements demonstrate markedly different usage patterns between malicious and undetected samples. The \texttt{com.apple.application-identifier} entitlement, which typically identifies an application through its bundle identifier prefixed with the team ID, appears in 17.9\% (8,029) of malicious samples compared to 7.8\% (3,485) of undetected samples. Similarly, \texttt{com.apple.security.get-task-allow}, which allows debuggers to attach and enables code injection, appears in 3.2\% (1,421) of malicious samples compared to 1.3\% (592) of undetected samples. Development tools like Xcode automatically add the entitlement to debug builds but remove it from release builds because it allows arbitrary code injection, posing a security risk. In contrast, entitlements like \texttt{com.apple.security.cs.allow-jit} appear more frequently in undetected samples at 14.2\% (6,293) compared to 0.8\% (379) in malicious samples, reflecting legitimate use cases in applications that require just-in-time compilation, such as browsers, virtual machines, and emulators.

Within our dataset, certain entitlements are overly present in the malicious binaries. 
  
\subsection{Certificates in macOS Malware} 
\label{mypar:certificates-macos-malware}

As our background section describes, code-signing certificates form a foundation for macOS security, establishing trust between developers and users by cryptographically verifying software authenticity. In the Apple ecosystem, certificates function as digital identities that tie binaries to their authors through unique Team IDs embedded in each certificate. After identity verification, developers obtain certificates through Apple's Developer Program, receiving credentials in the X.509 format that form part of a chain of trust anchored by Apple's Root Certificate Authority\cite{apple_tn3161_2024}. 

When developers sign code, the system embeds a secure timestamp within the signature, documenting when signing occurred, and allowing macOS to verify that the certificate remained valid at signing time even if the certificate later expires or becomes revoked. The timestamp mechanism creates a forensic trail for security researchers. When Apple discovers malware in the wild, it revokes associated developer certificates, creating a measurable timeline between initial code signing (development/deployment) and certificate revocation (discovery)\cite{apple_tn3161_2024}. By analyzing revocation timestamps, researchers can calculate the mean time to discovery for various malware families and campaigns, providing metrics about defensive effectiveness.

We analyzed 44,804 malicious macOS samples to study certificate usage and evasion strategies. Only 7.0\% (3,149 samples) of malicious binaries were signed, while the vast majority, 93.0\% (41,655), remained unsigned. Of the signed samples, 91.1\% (2,869 samples) used certificates from trusted certificate authorities (CA), accounting for 6.4\% of all malicious binaries, and 8.9\% (280 samples) used self-signed certificates, accounting for 0.6\% of all malicious binaries, which likely aim to simulate legitimacy in unmanaged environments.

\begin{figure}[t]
    \centering
    \includegraphics[width=0.4\textwidth]{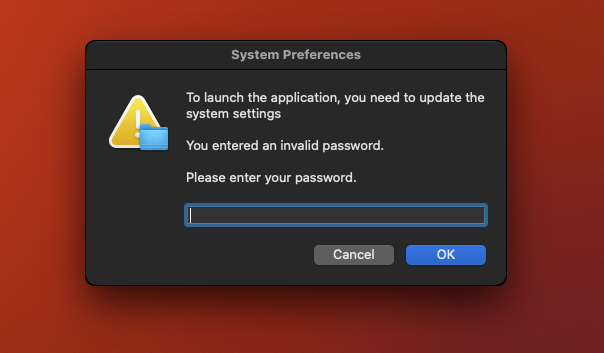}
    \caption{Example of social engineering in AMOS malware (MD5: 09be60c161ed0404b730fd906b178635). The malware displays a fake ``System Preferences'' dialog prompting users for their password under the pretense of updating system. The screenshot was generated by running the binary on a VMWare instance running on a mac mini.}
    \label{fig:amos_phishing_dialog}
\vspace{-0.5cm}
\end{figure}

Unsigned malware can execute via user-mediated override of Gatekeeper enforcement, typically induced through social engineering. macOS Malware (e.g., MD5: ee718b4b50d46a5c9fd5b94fc0b344d4) demonstrates this behavior by masquerading as legitimate software, such as a cracked application or installer, prompting users to manually approve execution. After launch, the malware presents a deceptive AppleScript dialog impersonating "System Preferences" that requests user credentials under the pretense of system configuration updates (Figure~\ref{fig:amos_phishing_dialog}). The malware validates stolen credentials using macOS Directory Services APIs and retries the prompt if validation fails, creating a persistent credential-harvesting loop until the user provides valid credentials.

AMOS additionally employs bypass techniques to evade Gatekeeper on subsequent executions. After initial execution, the malware copies itself to a hidden directory and programmatically removes the quarantine attribute \texttt{xattr -d com.apple.quarantine `\%s' $2\ge\&1$.}

Removing the quarantine attribute bypasses Gatekeeper inspection by eliminating the marker that identifies the file as downloaded from an untrusted source. The copied executable is then treated as locally created, allowing execution without triggering Gatekeeper checks. The \texttt{xattr} utility is a standard macOS tool for inspecting and modifying extended file attributes, including quarantine flags, which makes such a bypass technique difficult to prevent with system-level restrictions alone. AMOS then establishes persistence via LaunchAgent plists that automatically launch on user login, ensuring continued execution across system reboots.

Among CA-signed samples, 28.4\% (816 samples) had revoked certificates at the time of analysis. Several of these (e.g., MD5: 7aa48149d86b3c756e2bbe618f12929c and MD5: e258a2d3addcc9e2f2218b83d7bf838a) appeared on VirusTotal with multiple AV detections prior to revocation, suggesting that delays in certificate invalidation extend attackers' operational windows. Because detection metadata such as Team IDs and certificate identifiers was already visible to AV vendors at detection time, these results suggest that earlier certificate revocation would have been possible if similar intelligence had been incorporated into Apple's revocation pipeline. A tighter integration between AV reporting and Apple's certificate authority could therefore reduce the window during which compromised certificates remain valid.

Using AVClass2, we identified that a subset of the samples with revoked certificates belong to the \textit{Nukesped} malware family (31 samples, 3.8\%). While Nukesped is not among the most prevalent families in this subset, it is notable due to prior attribution by the United States Cybersecurity and Infrastructure Security Agency (CISA) and Palo Alto Networks Unit 42 to the Lazarus Group, a threat actor linked to the Democratic People's Republic of Korea (DPRK)~\cite{cisa_aa24_207a,cisa_2021_applejeus,unit42_2025_rustdoor_koi}. Across the Nukesped-labeled samples with revoked certificates, we observed repeated Apple Team IDs (e.g., "264HFWQH63", "2C4CB2P247") shared across multiple binaries, indicating coordinated reuse of signing identities within this cluster.

The Team ID is a unique alphanumeric identifier assigned by Apple to each registered developer account and included in every issued certificate. We independently extracted Team IDs from signing metadata to cluster binaries that share signing provenance. This identifier allowed us to correlate binaries across campaigns and to track certificate reuse by threat actors, a technique not previously documented. Further, the Team IDs enabled us to link revoked certificates to malware families and persistent infrastructure reuse.

We further extend our analysis by examining certificate revocation among malicious samples in our dataset. We analyzed 857 revoked CA-signed certificates associated with malicious binaries to characterize revocation timing. The mean time from certificate issuance to revocation is 161 days, computed as: $$\text{Mean Time-to-Revocation} = \frac{1}{N}\sum_{i=1}^{N}(t_{revocation_i} - t_{issuance_i})$$ where $N$ is the total number of certificates analyzed, $t_{revocation_i}$ is the timestamp when Apple revoked the $i$-th certificate and $t_{issuance_i}$ is the date the certificate was issued.
This average masks substantial dispersion. The median time to revocation is 1 day, while the 90th percentile reaches 576 days, and several certificates remained valid for multiple years. The largest observed interval is 4{,}367 days (approximately 12 years) for a certificate issued to TeamID \textit{94KSDA266A} (``TMRG, Inc.'') in February 2013 and revoked in January 2025. The first associated sample appeared on VirusTotal in May 2015, indicating that the certificate remained unrevoked for 3{,}537 days (9.7 years) after public detection.

A notable example is the certificate issued to TeamID \textit{AGN79H7MTU} (``Ruth Hall''), which remained valid for 721 days (nearly 2 years) before revocation. The certificate was issued in December 2022 and remained in effect until December 2024, despite the malware being associated with the Lazarus Group, a well-documented DPRK threat actor. The malware operated undetected for 255 days before a sample (MD5: b91729abc87fffbcc1be075ec57307e1) first appeared in VirusTotal in September 2023. Remarkably, despite being publicly available in VirusTotal, the sample remained undetected by antivirus engines for over 15 months until ESET NOD32 first flagged it as malicious on December 17, 2024, one week after the certificate was finally revoked on December 10, 2024. This ordering shows that certificate revocation in this case did not follow initial antivirus detection and highlights an opportunity to reduce revocation latency within Apple's certificate revocation infrastructure. More broadly, it suggests the value of tighter coordination between certificate authorities and security vendors to limit reuse of compromised signing identities.

\subsection{Notarization in macOS Malware}
Apple’s notarization service, introduced in macOS 10.14.5, requires developers to submit applications to Apple for automated security screening prior to distribution. 

\textbf{Notarization of Known Malware.}
Stealers provide a concrete lens into notarization latency. We observe cases where samples with extensive antivirus consensus later acquired notarization records. For example, a credential-stealing sample (MD5:a8b54bc4b2d3723b11666fc5fe646cdb) demonstrates the temporal gap between macOS notarization and malware detection. Katalina shows that a notarization ticket was issued for the binary’s CDHash (be2ae0a004f0a8803a5b576855697bed2df3b91f) on 2025-07-15T22:23:09Z, indicating that the sample successfully passed Apple’s notarization checks at that time. The binary first appeared on VirusTotal on 2025-08-26T07:14:15Z, with the signing certificate (TeamID \emph{3GUHMVK4XV}, Subject Organization \emph{Alina Balaban}) revoked later the same day at 2025-08-26T19:00:00Z, and by September~23,~2025, 19 antivirus engines had identified it as malicious. While no notarization ticket revocation is observable for this binary, certificate revocation alone is sufficient for Gatekeeper to block execution on systems enforcing current trust checks; however, the key distinction is that the malware was successfully notarized, indicating it was not identified as malicious during the notarization process.

While these cases do not reveal the internal reasons for approval, they demonstrate that notarization decisions are not tightly synchronized with external malware detection signals, motivating closer study of cross-signal integration between detection ecosystems and notarization workflows.


\textbf{Implications for Malware Distribution.}
The presence of notarization can provide operational advantages to attackers. When malware is both properly signed and notarized, macOS Gatekeeper typically allows it to execute without prominent warnings, which may lead users to perceive the software as trustworthy despite its malicious behavior. The presence of notarization among malicious samples indicates that automated screening at submission time (pre-distribution) does not reliably detect malicious behavior under adversarial conditions. The concentration of notarized samples within specific families, such as AMOS, further suggests that some malware ecosystems repeatedly obtain notarization.

\subsection{Abuse of Native Scripting Interfaces}
\mypar{osascript}
On macOS, \texttt{osascript} executes AppleScript and other Open Scripting Architecture (OSA) scripts. Although widely used in automation and device management (e.g., Jamf), its dual-use nature mirrors tools like PowerShell and Bash, commonly abused in Windows malware.

We identified 1,628 (3.63\%) malicious samples in MALET that invoke \texttt{osascript}, often through the -e flag, which allows inline execution of AppleScript commands. AVClass2 assigned many of these samples to unknown generic labels such as  \texttt{UNK:amos} and \texttt{UNK:banshee}. Through manual spot-checking of 100 samples, we confirmed that these labels correspond to malware families, including \textit{AMOS} and \textit{Banshee}. These families use \texttt{osascript} for phishing and data exfiltration. A typical tactic involves spoofing native UI dialogs to capture passwords. For example, AMOS uses:

\begin{lstlisting}[
    frame=single,
    framerule=1pt,
    language=bash,
    basicstyle=\footnotesize\ttfamily,
    breaklines=true,
    columns=fullflexible,
    showspaces=false,
        showstringspaces=false,
]
osascript -e 'display dialog "MacOS wants to access System Preferences" \
    with title "System Preferences" with icon caution \
    default answer "" with hidden answer'
\end{lstlisting}
Banshee uses similar dialogs, often verifying input via \texttt{dscl} which is a directory service utility used to query local user accounts before exfiltrating credentials\cite{dscl}. Unlike Windows and Linux malware, which often harvest secrets from memory, these macOS families rely on deceptive user interaction.

Beyond social engineering, \texttt{osascript} supports file discovery and interaction with the app. AMOS and Banshee scripts target files with extensions like \texttt{.wallet}, \texttt{.txt}, and \texttt{.key}, automating their collection from common directories, paralleling the behavior of PowerShell and Bash on Windows and Linux correspondingly.

\subsection{Use of Packers in macOS Malware}
\label{mypar:packers-macos-malware}

While packers are prevalent in Windows malware, their use in macOS appears limited. Our static analysis of the MALET dataset identified 610 malware samples that were packed, representing 1.36\% of the total malware in MALET. We identified Ultimate Packer for Executables (UPX) \cite{upx} as a consistently used packer in families such as Sliver, Cointhief, Vshell, and macOS builds of Cobalt Strike. No other advanced known packers, such as Themida, MPress, or VMProtect, were detected

Packing was rare in undetected samples, with only 6 samples (0.01\%) showing packer signatures. This suggests that the presence of a packer may serve as a low-noise heuristic for identifying potentially malicious samples. Though UPX is not inherently malicious, its exclusive appearance in malware within our dataset highlights its operational role. Compared to Windows, the macOS ecosystem shows a much lower prevalence of complex packers, i.e, Themida, MPress, or VMProtect. In Windows, sophisticated, layered protectors are common. Prior work on Linux malware~\cite{cozzi2018understanding} suggests an intermediate position, with widespread use of UPX and occasional deployment of custom packers. Some Linux samples include modified headers and unpacking stubs to obscure analysis, a pattern not observed in our macOS dataset. 

It is important to acknowledge that our observations might be influenced by the limitations of current detection capabilities or our static analysis methodology. Nonetheless, the predictable nature of UPX presents an opportunity for further exploration. We leave a deeper investigation as future work.

\section{macOS Malware Classification} \label{section:classification-macos-malware}
\label{sec:macos_malware_classification}
macOS malware classification remains underdeveloped and hindered by inconsistent Anti-Virus (AV) labels, limited detection accuracy, and poor taxonomy adoption. Unlike Windows and Linux, which benefit from mature family definitions, macOS classification lacks consistency, complicating attribution and threat modeling~\cite{cozzi2018understanding}. In this work, we use AVClass2~\cite{avclass2}, a publicly available tool, to extract family and class labels from the output of the AV engine. Although AVClass2 enables reproducible classification from noisy scan data, its effectiveness is ultimately limited by the quality of the underlying AV labels.

\subsection{AV Engine Labeling Problems}
AV engines frequently mislabel credential-stealing malware as belonging to the \texttt{AMOS} family. This labeling pattern appears to arise from static rule sets that focus on specific traits, such as references to the macOS keychain or browser credential access, without accounting for deeper structural context. One clear example is a distinct macOS stealer cluster we identified, referred as \texttt{Banshee}. Despite its unique codebase and execution model, AV engines overwhelmingly applied the \texttt{AMOS} label because the binaries target browser credentials, including Chrome, Edge, and Firefox.

To validate the distinction between \texttt{Banshee} and \texttt{AMOS}, we performed manual reverse engineering in Ghidra. We compared symbolic information from the MALET dataset with decompiled method bodies to assess semantic behavior. Our analysis confirmed the presence of Objective-C class names such as \texttt{AntiVM}, \texttt{Wallets}, and \texttt{Browsers}, none of which collectively appear in verified \texttt{AMOS} samples. We also found the use of other unique methods among Banshee samples, such as \texttt{isRussianLanguageInstalled} that checks the system's language settings, likely to avoid infecting Russian-speaking users, a tactic observed in malware developed by Russian cybercriminals to evade domestic law enforcement scrutiny\cite{chainalysis2021eastern}. Other methods, including \texttt{collectDataAndSave}, \texttt{pathToProfile}, and \texttt{pathToSave}, coordinate the exfiltration of browser data. Functions such as \texttt{setChromePath} and \texttt{collectWalletData} indicate browser-specific and wallet-focused credential theft workflows.

A representative example is the binary with MD5 hash \texttt{a6cfd4597855fb766348731e081a4edb}. AV engines labeled it as \texttt{AMOS}, and AVClass2 reflected that consensus by assigning the label \texttt{UNK:amos} as the default family for any stealer with credential theft capabilities.
In total, we identified 85 mislabeled samples within the \texttt{Banshee} cluster. Notably, 46 of these shared the same \texttt{ObjcMethodHash}: \texttt{0cdc6580ec2760151a14d41a6ee66680f4291e6e92\\345783197f6ba6e79a0736} or \texttt{ObjcClassHash}: \texttt{4ea964155b68f1981071ce907ff0482046c9ff90c5f\\a8887d1bd4ecd69de8dc2}. This value represents a normalized hash of Objective-C class names extracted from each binary using our static analysis framework, Katalina. The \texttt{ObjcClassHash} serves as a structural fingerprint that captures the underlying class composition, which remains consistent despite minor variations in the binary. The misclassification pattern reflects a broader issue in macOS malware labeling. When AV engines apply narrowly scoped static rules that fail to incorporate semantic and structural variation, they group unrelated binaries under the same label. As a result, malware with distinct implementations and attribution paths remains incorrectly categorized, weakening the accuracy of threat intelligence.

\subsection{AVClass2 Limitations}
\label{section:avclass2}

AVClass2 is an open-source tool that extracts structured family, class, and behavior tags from noisy AV engine outputs. It uses token co-occurrence analysis and rule-based expansion to standardize disparate vendor labels, enabling large-scale malware classification and analysis. While its taxonomy is well-developed for Windows and Android samples, macOS coverage is less mature, which can affect classification outcomes for this platform

In the MALET dataset, AVClass2 did not assign family labels to 24,763 samples (55.27\%), class labels to 16,897 (37.71\%), and behavior tags to 10,563 (23.58\%). Additionally, 8,011 samples (17.88\%) lacked all classification tags (e.g., \texttt{FAM:}, \texttt{CLASS:}, \texttt{BEH:}). Even when labels were extracted, 22,160 (49.46\%) samples had multiple candidate class tags and 18,967 (42.33\%) had multiple candidate behaviors, reflecting ambiguity in AV engine descriptions.

AVClass2 uses token co-occurrence and rule-based expansion to identify the most likely tags for each sample. When multiple family tags are present, it selects a single top-ranked label, which can suppress alternative possibilities. In some cases, this may obscure nuances in the data or reinforce a majority view, as seen in the \texttt{AMOS}-\texttt{Banshee} misclassifications. Its default taxonomy omits several families active in macOS campaigns, leading to placeholder labels such as \texttt{UNK} or assignment to unrelated Windows or Android families. Table~\ref{tab:classification-summary} summarizes these observations, which highlight opportunities to strengthen macOS coverage in automated malware classification.

    

\begin{table}[t]
\centering
\caption{A distribution of our observed misclassification across family, class, and behavior.}
\begin{tabularx}{\linewidth}{X|r|r}
\toprule
\textbf{Analysis} & \textbf{Count} & \textbf{Percentage (\%)}\\
\midrule
Missing Family Label              & 24,763 & 55.27 \\
Missing Class Label               & 16,897 & 37.71 \\
Missing Behavior Label            & 10,563 & 23.58 \\
Families with Multiple Classes    &     56 &  0.13 \\
Multiple Candidate Classes        & 22,160 & 49.46 \\
Multiple Candidate Families       &  2,028 &  4.53 \\
Multiple Candidate Behaviors      & 18,967 & 42.33 \\
No Classification Tags            &  8,011 & 17.88 \\
\bottomrule
\end{tabularx}
\label{tab:classification-summary}
\end{table}

\section{Further Observations}
Building upon our analysis of malware classification challenges, we conducted follow-up investigations to assess the real-world impact of classification inconsistencies on threat detection and explore the potential of advanced techniques to improve classification accuracy. These observations provide a practical context for the theoretical limitations identified earlier and offer insight into promising directions to address the classification challenges unique to macOS malware.

\subsection{Misclassified Samples with 0 Detections}
\label{mypar:uuid-collisions}
As shown in Phase~3 of Figure~\ref{fig:malet_processing}, we identified samples with zero antivirus detections (\emph{undetected}) that did not satisfy our \emph{high-confidence benign} criteria. We clustered these zero-detection samples together with known malicious binaries from MALET using a set of five structural features that together form a stable code fingerprint: LC\_UUID (identifying unique builds~\cite{apple_tn3178}), a hash of externally linked symbols (ExtSymHash), a hash of all symbol names (AllSymHash), a hash of Objective-C method names (MethodHash), and a hash of Objective-C class names (ClassHash). These features persist across repackaging and reflect intrinsic code structure rather than signing identity or superficial metadata.

Using this multi-feature clustering approach, we identified \textbf{145 clusters} of binaries that share identical structural fingerprints despite having different detection outcomes. In total, these clusters contain \textbf{3,134 binaries}, including \textbf{2,890 undetected samples} (zero AV detections) and \textbf{247 labeled as malicious}. The number and size of these clusters indicate systematic code reuse across campaigns rather than incidental repackaging, and suggest that a non-trivial fraction of malware remains undetected despite substantial code overlap with known malicious samples.

Specifically, we analyzed a cluster containing samples signed with certificate serial \texttt{187aa9a8c296210c} issued to ``THOMAS BOULAY DUVAL'' with TeamID \texttt{J97GLQ5KW9}. These samples were distributed through social engineering campaigns where threat actors impersonated Google recruiters to deliver malicious payloads disguised as legitimate applications~\cite{cybernewsev2025}. Among these, one sample (MD5 \texttt{7f49e4d8d658c0b2535b0f605f0711ea}) was flagged as malicious with 14 AV detections, while a structurally identical sample (MD5 \texttt{fcf35ef387f9175cbd1806e5ff718529}) received zero detections from any AV engine. Both samples share identical architecture-independent features: bundle identifier (\texttt{thomas.parfums}), entitlement hash (\texttt{10f14189818...}), Swift type hash (\texttt{944ce0ed449...}), dylib dependencies (24 libraries including SwiftUI and Combine frameworks), and C2 domain (\texttt{franceparfumes.org}). The Swift type hash confirms that both samples implement the same application structure, including \texttt{CartView}, \texttt{CheckoutView}, \texttt{PerfumeStoreApp}, and \texttt{ProductCard}. The only difference between samples is CPU architecture: the detected sample targets ARM64 while the undetected sample targets x86\_64.

The cluster demonstrates a fundamental weakness in current macOS malware detection. Current detection practices often fail to fully exploit the generalizable structural features inherent to Mach-O binaries. Features such as LC\_UUID, symbol information, code-signing properties, and notarization metadata persist across campaigns and architectures, yet are not consistently leveraged by detection systems. Notably, the five features employed in this analysis represent only a small subset of available structural characteristics in Mach-O binaries. Our dataset contains additional clusterable attributes including certificate chain properties, dylib dependencies, segment and section layouts, code directory hashes, notarization metadata, instruction-level patterns, and string artifacts. Despite this, structurally identical malware can receive inconsistent detection outcomes based on superficial variations rather than substantive code differences.

To illustrate the effect of structural collisions in practice, in Table~\ref{tab:feature-collision-summary}, we present example binaries drawn from a cluster of structurally identical samples that nevertheless received different labels.

\begin{table}[!t]
\centering
\caption{Example samples from a structurally identical cluster. All binaries share the same certificate serial (\texttt{187aa9a8c296210c}), TeamID (\texttt{J97GLQ5KW9}), bundle identifier (\texttt{thomas.parfums}), entitlement hash, Swift type hash, dylib dependencies, and C2 domain, but differ in detection outcomes.}
\label{tab:feature-collision-summary}
\small
\begin{tabular}{p{7cm}|l}
\toprule
\textbf{Sample MD5} & \textbf{Label} \\
\midrule
fcf35ef387f9175cbd1806e5ff718529 & Benign \\
7f49e4d8d658c0b2535b0f605f0711ea & Malicious \\
\bottomrule
\end{tabular}
\end{table}

\section{Threats to Validity}
\label{sec:limitation}

\mypar{Lack of Raw Binaries}
Our MALET dataset does not provide the raw malware binaries. This intentional absence avoids the ethical and legal complications of distributing live malware. However, this absence introduces limitations. Specifically, models and tools that require byte-level input cannot operate directly on our dataset. To support reproducibility and further research, we include both the SHA256 file hashes and the Katalina source code, enabling researchers with appropriate access to retrieve the binaries independently.

\mypar{Static Analysis Limitations}
Our behavioral insights are inferred solely from static features extracted from each binary. As a result, we cannot observe runtime behavior, dynamic unpacking routines, or command-and-control communications that only manifest during execution. This limitation prevents us from comprehensively analyzing malware functionality that relies on time-based activation, encrypted payloads, or other techniques evading static inspection.

\mypar{Dataset Bias}
Our malware samples were sourced from VirusTotal and VirusShare, potentially biasing the dataset toward threats that are more easily detected or shared with vendors. Consequently, insights from MALET may not generalize to proprietary or undetected malware. Benign samples, drawn mainly from macOS system files, may not capture the full diversity of real-world applications. The dataset also reflects a fixed point in time and may not represent future shifts in attacker behavior.

\section{Related Work}
\label{sec:related_work}
Malware research has historically concentrated on non-macOS platforms, with macOS receiving less attention despite growing enterprise adoption. This section reviews malware datasets and analysis, macOS malware analysis approaches, and relevant classification techniques.

\mypar{Dataset} The Android Malware Genome Project~\cite{zhou2012dissecting}, AMD~\cite{wei2017deep}, and EMBER~\cite{anderson2018ember} established foundational datasets for Android and Windows malware research. For Linux, Cozzi et al.~\cite{cozzi2018understanding} analyzed 10,548 samples, finding most were unsophisticated crypto miners or DDoS bots. In contrast, macOS research lacks comprehensive datasets, with Reed et al.~\cite{reed2020survey} noting this gap. 

\mypar{Malware analysis}Malware analysis encompasses a diverse set of techniques aimed at understanding adversarial behavior and designing effective defenses~\cite{bilge2011exposure, blazytko2017syntia, canzanese2015runtime, christodorescu2007mining, comparetti2010dormant, cui2012practical, dinaburg2008ether, hassen2017static, jacob2011jackstraws, jagpal2015extensions, kim2017dynodet, kirat2014malgene, kirat2014barecloud, kolbitsch2010gadget, kolbitsch2011procrastination, martignoni2008layered, palahan2013significant, peng2014xforce, rieck2008learning, tegeler2012botfinder, xu2017platpal, yen2008traffic}. Research efforts span static analysis, which examines binaries without execution, and dynamic analysis, which observes behavior at runtime. Hybrid methods combine both for improved visibility into malware functionality. Innovations such as hardware-assisted sandboxes~\cite{dinaburg2008ether}, memory introspection~\cite{cui2012practical}, and statistical behavior mining~\cite{palahan2013significant, rieck2008learning} have significantly expanded detection capabilities. However, challenges persist, especially with obfuscated code~\cite{blazytko2017syntia}, dynamic evasion~\cite{kirat2014malgene}, and platform-specific behavior. Our work builds on these foundations by applying them to the underexplored macOS landscape, incorporating static structural analysis into a dataset of unprecedented scale and scope.

\mypar{macOS malware analysis}
macOS malware analysis has evolved significantly over the years, addressing the unique challenges posed by Apple's operating system. Early research by Wardle~\cite{wardle2017methods} provided a comprehensive overview of macOS malware techniques, focusing on persistence mechanisms and highlighting the need for specialized analysis tools. Building upon this, Pham et al.~\cite{pham2019mac} introduced Mac-A-Mal, a framework designed to resist anti-evasion techniques commonly employed by macOS malware. Their approach utilized a kernel extension to monitor malware behavior, effectively mitigating several anti-evasion strategies.
In ransomware, Kolodenker et al.~\cite{kolodenker2017ransomware} analyzed macOS ransomware, emphasizing the importance of proactive detection and response strategies. Their work underscored the necessity of understanding the unique behaviors and characteristics of macOS-targeted ransomware.
Entitlements in macOS, akin to Android permissions, play a crucial role in platform security. While Android permissions have been extensively studied~\cite{felt2012android, au2012pscout}, macOS entitlements remain relatively understudied. Wang and Wu~\cite{wang2018exposing} demonstrated API abuse across Apple platforms, highlighting the privacy implications of entitlement misuse and the need for more in-depth analysis in this area.
Static analysis of macOS binaries presents challenges due to the Mach-O format and platform-specific metadata. Levin~\cite{levin2021apple} provided detailed documentation on the Mach-O format, facilitating a better understanding and analysis of macOS binaries. Egele et al.\cite{egele2011piosecurity} developed techniques for analyzing Apple binaries, contributing to the advancement of static analysis methodologies. Zhu et al.\cite{zhu2022understanding} examined the implications of architectural transitions, such as the shift to Apple Silicon on malware analysis and detection.
Apple's security mechanisms, including System Integrity Protection (SIP), have been subjects of study by multiple researchers. Levin~\cite{levin2021apple} analyzed SIP, providing insights into its role in protecting system files and processes. Wang and Wu~\cite{wang2018exposing} demonstrated how applications could bypass such restrictions, emphasizing the need for continuous evaluation and enhancement of macOS security features.

Our work differs from previous efforts in four key aspects: 
\begin{itemize}
    \item To the best of our knowledge, MALET provides an extensive public dataset of macOS malware with standardized features;
    \item Katalina extracts platform-specific attributes previous tools overlooked;
    \item Our characterization reveals macOS-specific behaviors and detection gaps;
    \item We establish foundations for cross-platform comparative security research contextualizing macOS threats.
\end{itemize}
\section{Future Work}
 \label{sec:future_work}
MALET and Katalina establish a foundation for reproducible, large-scale analysis of macOS malware. However, several unexplored directions can extend the utility of our dataset and toolchain.

\mypar{Beyond Labels: Toward Semantic macOS Malware Classification}
 AVClass2 and antivirus engines often fail to distinguish macOS malware families that differ structurally and behaviorally. Section~\ref{section:classification-macos-malware} outlines mislabeling patterns, while Section~\ref{mypar:uuid-collisions} shows that binaries with identical structures receive conflicting verdicts. These findings point to two future directions: improving malware labeling and advancing malware detection.
 Labeling systems should incorporate macOS-specific traits and expand the corpus used by tools such as AVClass2. The new taxonomies must reflect the behaviors and families of macOS. Researchers can apply large language models to increase label coverage by reasoning over structural features and metadata extracted by Katalina.
Detection tools should use transparent and reproducible MALET features. Static signals such as load command patterns, entitlement sets, section entropy, and instruction distributions support lightweight classifiers. Future work applying machine learning should follow the best practices outlined by Arp et al.~\cite{arp2020and} to minimize common pitfalls in applying machine learning to malware detection.
Progress in both labeling and detection will enable  robust and interpretable macOS malware classifiers. Finally, the research community could use the malicious parts of the MALET dataset and contribute Sigma rules based on the patterns of malicious behaviors observed at scale.

\mypar{Investigating Entitlements for Risk Modeling and Detection}
 Entitlement misuse remains one of the clearest markers of malicious behavior in macOS malware. Section~\ref{mypar:entitlements-macos-malware} shows how malware often requests fewer but riskier entitlements than benign software. Inspired by previous Android research on the cooccurrence and abuse of permissions~\cite{au2012pscout,alam2017droidnative}, future work should model entitlement interactions to detect anomalous or high-risk combinations. The tools can score binaries based on the rarity of entitlements, the scope of privileges, and the historical abuse patterns observed in MALET. These scores can support lightweight malware detectors or help triage binaries for manual review, similar to how Android permissions inform APK risk analysis in apps such as Andrubis~\cite{lindorfer2014andrubis}.

\mypar{Longitudinal Analysis of Certificates and TeamIDs}
 Section~\ref{mypar:certificates-macos-malware} details how attackers, including APT actors, reuse Apple Developer IDs and exploit slow revocation. Future work can trace TeamID reuse, certificate abuse, and time-to-revocation metrics on broader timescales. These analyses can help identify persistent infrastructure, improve the early detection of signed malware, and guide stricter issuance and monitoring policies. Combining certificate timelines with behavioral features may also improve threat actor attribution.

\mypar{Detecting packers and obstruction in macOS Malware}
 Section~\ref{mypar:packers-macos-malware} reveals that while packers like UPX are rare in benign software, they appear in malware samples across several families. Future work should develop classifiers that detect packing and obfuscation using static features in MALET, such as entropy distributions and assembly instruction profiles. These features already exist in the dataset and can be used to build simple ML-based classifiers or rule-based heuristics. Similarly to previous work on Linux malware~\cite{cozzi2018understanding}, these detectors can help flag evasive samples or identify custom packing routines in future macOS threats.


\section{Conclusion}
\label{sec:conclusion}

MALET is a large-scale, publicly available dataset for macOS malware research, comprising 44{,}804 malicious and 44{,}457 benign Mach-O binaries with rich, structured static features. To support this corpus, we release Katalina, an open-source framework that extracts macOS-specific attributes and enables reproducible, large-scale empirical analysis. Using MALET, we surface multiple detection blind spots, including the continued prevalence of unsigned malware, systematic misuse of native scripting capabilities, and inconsistent antivirus labeling across structurally identical binaries.

Our analysis further highlights weaknesses in code-signing trust infrastructure in practice. Although Gatekeeper incorporates multiple policy checks, unsigned malware may still execute through user-mediated approval, and a substantial fraction of signed malware persists until certificate revocation. Among CA-signed malicious samples, 28.4\% were associated with revoked certificates, with revocation often occurring long after public detection. We also observe cases of certificates linked to publicly attributed threat groups, including DPRK-associated actors, as reflected in antivirus labels and threat intelligence reports. In parallel, labeling noise remains substantial, with nearly one third of malware samples receiving no decisive family classification.

Together, MALET and Katalina provide a foundation for more transparent evaluation, improved attribution research, and macOS-specific detection engineering. We believe that open measurement of malware ecosystems at scale is essential to strengthening platform security and enabling defensible, data-driven improvements in malware detection and response.



\bibliographystyle{IEEEtran}
\bibliography{malet-ref}


\end{document}